\documentclass{article}

    \PassOptionsToPackage{numbers, compress}{natbib}

    \usepackage[preprint]{neurips_2024}

\usepackage[utf8]{inputenc} 
\usepackage[T1]{fontenc}    
\usepackage{hyperref}       
\usepackage{url}            
\usepackage{booktabs}       
\usepackage{amsfonts}       
\usepackage{nicefrac}       
\usepackage{microtype}      
\usepackage{xcolor}         

\usepackage{graphicx}

\title{What AI evaluations for preventing catastrophic risks can and cannot do}

\author{%
  {Peter Barnett \hspace{2cm }Lisa Thiergart}\\
  Machine Intelligence Research Institute\\
  \texttt{\{peter, \ lisa\}@intelligence.org} 
}

\begin{document}

\maketitle

\begin{abstract}
AI evaluations are an important component of the AI governance toolkit, underlying current approaches to safety cases for preventing catastrophic risks. Our paper examines what these evaluations can and cannot tell us. Evaluations can establish lower bounds on AI capabilities and assess certain misuse risks given sufficient effort from evaluators.

Unfortunately, evaluations face fundamental limitations that cannot be overcome within the current paradigm. These include an inability to establish upper bounds on capabilities, reliably forecast future model capabilities, or robustly assess risks from autonomous AI systems. This means that while evaluations are valuable tools, we should not rely on them as our main way of ensuring AI systems are safe. We conclude with recommendations for incremental improvements to frontier AI safety, while acknowledging these fundamental limitations remain unsolved.

\end{abstract}

\section{Introduction} 
In our previous paper, "Declare and Justify: Explicit Assumptions in AI Evaluations are Necessary for Effective Regulation," we analyzed the assumptions underlying AI evaluation frameworks and argued that as part of a safety case these should be explicitly stated and justified~\cite{barnett2024declare}. This paper builds on that analysis to examine what AI capability evaluations can and cannot accomplish in preventing catastrophic risks, with particular attention to their role in governance decisions. 

AI evaluations play a crucial role in safety cases—structured arguments that an AI system is safe~\cite{clymer_safety_2024, goemans2024safety, anthropic2023responsible, openAI2023preparednes, deepmind2024frontier}. These safety cases often rely on two key premises: first, that if a model does not demonstrate a capability in evaluations, it likely cannot cause harm using that capability; and second, that before AI systems develop dangerous capabilities, evaluators will observe precursor capabilities. However, the validity of these premises depends heavily on the underlying assumptions about how AI capabilities can be measured and predicted.

The challenges and limitations of AI evaluations can be better understood by considering two key dimensions:
\begin{enumerate}
    \item \textbf{Timing}: Separating the evaluation of existing models from the forecasting of future model capabilities. The key difference is that existing models can be directly interacted with, while future model capabilities must be indirectly inferred from existing models.
    \item \textbf{Risk type}: Differentiating between risks from human misuse and risks from AI systems acting autonomously in misaligned ways. Risks from human misuse may be easier to evaluate than risks from misalignment, which may prove intractable.
\end{enumerate}
We discuss what AI evaluations can be used for, including what they could do with significantly more effort from evaluators, and what fundamentally cannot be done under the current paradigm. Our analysis reveals both the useful role evaluations can play in AI safety, and their inherent limitations, with important implications for AI governance and regulation.

\section{What AI evaluations can do (given sufficient effort)}

\subsection{Establish lower bounds on capabilities}
Evaluations can provide concrete evidence of what AI systems are capable of doing~\cite{phuong_evaluating_2024}. When an AI system successfully completes a task—whether identifying a class of cybersecurity vulnerabilities~\cite{fang2024llmwebsites, fang_llm_2024, fang_teams_2024}, or manipulating human users in a controlled environment~\cite{salvi_conversational_2024, matz_potential_2024}—this conclusively establishes the presence of those capabilities. These demonstrated capabilities form a lower bound: we know for certain the AI system can do at least this much.

However, it is not clear how much a capability displayed in one context will translate into another.  When evaluating similar tasks under controlled conditions, we can make statistical predictions about tasks drawn from the same set. For instance, if an AI system correctly answers 60\% of cybersecurity questions from a standardized test set, we can predict similar performance on other questions from that same set. However, in real-world scenarios, tasks are not drawn from standardized test sets. Evaluations rely on the assumption that the measured tasks are a good proxy for the real-world tasks. Real-world applications may enable significantly higher performance than what we observe in controlled evaluations. Our proxy tasks may be too narrow or unconstrained settings may allow for novel approaches evaluators didn't test for.

These lower bounds provide useful but not sufficient guidance for safety requirements. Security measures must at minimum guard against demonstrated capabilities. Because this is a lower bound, we know our protective measures need to go beyond these lower bounds, but we cannot be certain how far beyond. 

\subsection{Assess misuse risk for current models}
In principle, evaluations may be useful for assessing the misuse risk for existing AI systems. This requires evaluation teams to outperform potential attackers in identifying and exploiting threat vectors. There is a clear bar: evaluators must be capable of identifying and reliably addressing all threat vectors that malicious actors could exploit. Under this approach, an AI system is assumed not to increase misuse risk if evaluators are unable to use it to exploit any threat vectors.

For misuse evaluations to work, evaluators must be better than attackers at finding dangerous capabilities. This may require that attackers do not have access to certain relevant models and tools, in order to give evaluators an advantage. Meeting these conditions in practice is extremely challenging. For example, if model weights are open-sourced or leaked, attackers will be on essentially the same footing as the evaluators. 

While meeting these conditions may be feasible \emph{in principle}, this is not the case with current evaluations. It is extremely likely that current evaluations will fail to consider important threat vectors or under-elicit AI system capabilities. Evaluations would then fail to find latent dangerous AI capabilities. Evaluators must be confident that they are not missing capabilities which malicious actors could exploit. 

This may be more justifiable if evaluators have affordances which malicious actors would not. For example, evaluators may have access to models which have not been trained for safety, or access to fine-tuning in order to train specifically for dangerous capabilities~\cite{casper_black-box_2024}. This advantage requires robust security measures to prevent unauthorized access to model weights. If a safety case depends on evaluator-only affordances, then preventing theft or leakage of model weights  is critical. Experts believe that frontier model weights could currently be stolen by determined state-actors~\cite{nevo2024securing}, or may already have been~\cite{venturebeatMistralConfirms}. The security of companies developing frontier AI is far from being robust to these attacks. 

Stopping more sophisticated malicious actors requires more thorough evaluations. Evaluators can likely outperform amateur actors who lack resources and expertise. However, evaluators might not match the capabilities of nation-state actors who have significant resources and specialized expertise. A further difficulty lies in assessing the maximum capability level of such a nation state actor, since this information is classified and inferences based on public information are limited in scope. 

For AI systems that don't significantly exceed human capabilities, the concern about sophisticated malicious actors may be less pressing. Such actors might find it easier to pursue their goals directly rather than invest significant resources in eliciting AI capabilities. But if AI systems exceed human capabilities or enable dangerous automation at scale, the situation changes. If an AI system offers a malicious actor novel capabilities or cost savings that make the effort worthwhile, they may be willing to invest heavily in capability elicitation. This also applies to AI systems that perform below human level, if they could enable unprecedented attacks through automation.

It may become easier to elicit capabilities of future AIs, especially for agentic tasks. These systems may be specifically trained for agentic tasks or otherwise be natively competent at these tasks. They may rely less on scaffolding and specialized tools, which today's AI systems require. But this is still a very uncertain prediction.

\subsection{Applications that don’t directly mitigate catastrophic risks}
Evaluations can serve important functions that support safety efforts and help society prepare for advanced AI systems.

First, evaluations contribute to the \textbf{basic science of AI}, improving our understanding of these systems. They can reveal how capabilities tend to develop under the current paradigm and help us understand how factors like architecture, model size, training data, and algorithms affect model behavior. Additionally, conducting evaluations advances the science of AI evaluation itself, helping develop better capability elicitation methods and reducing the risk of missing important capabilities.

Second, evaluations can help anticipate and \textbf{prepare for AI impacts on society}. They can provide early warning about jobs likely to be automated, demonstrate potential misuse cases (even if the evaluated system itself isn't misused), and highlight concerning concentrations of power that may result as private or state actors develop increasingly capable AI systems.

Finally, evaluations can serve as \textbf{coordination points for governance decisions}~\cite{alaga2023coordinated}. They provide concrete evidence to inform policy discussions, establishing shared reference points for debates about AI regulation and development pace. When evaluation results trigger safety concerns, this can help motivate and build support for implementing specific safety requirements, additional oversight measures, or restrictions on certain types of AI development until arguments for safety can be more rigorously justified (which prove to be very challenging as discussed below).
\section{What AI evaluations cannot do}
The following limitations represent fundamental challenges. Within the current paradigm, they cannot be overcome simply via more thorough evaluation.

\subsection{Establish upper bounds on capabilities}
Evaluations cannot establish upper bounds on AI system capabilities, because they do not provide strong evidence about the lack of capabilities. There are no principled methods to tell whether capabilities are being optimally elicited. Further training, additional fine-tuning and scaffolding, post-training enhancements, and novel interaction methods could result in a model succeeding at tasks it previously failed. 

The problem of under-elicitation is well-demonstrated by recent cybersecurity evaluations. \mbox{CyberSecEval 2}, a benchmark for discovering and exploiting vulnerabilities, initially showed GPT-4 Turbo exploiting only 5\% of buffer overflow vulnerabilities~\cite{bhatt2024cyberseceval}. However, when Project Naptime introduced a new scaffolding approach~\cite{googleprojectzeroProjectNaptime}, the same model could exploit 71\% of these vulnerabilities after a single attempt, and 100\% when allowed multiple attempts. Similar dramatic improvements have been achieved through various post-training enhancements, enabling models to perform at the level of much larger systems~\cite{davidson2023ai}. Even benchmarks which have not seen sudden jumps, often see steady but significant improvement when run with the same underlying AI model~\cite{jimenez2024swebench, swebenchSWEbench,huggingfaceGAIALeaderboard,paperswithcodeHumanEval}. 

The challenge of establishing upper bounds is further complicated by in-context learning~\cite{dong2024survey}. An AI system might demonstrate partial competence at several component skills of a dangerous capability. In such cases, it’s difficult to predict whether the system could acquire missing components through in-context learning. Similarly, a system that barely fails at all component tasks might appear safe, yet could plausibly reach the required skill level through in-context learning. These issues fundamentally undermine attempts to demonstrate safety through the absence of capabilities. When an evaluation fails to elicit a capability, this does not on its own provide strong evidence that the system lacks this capability.

This limitation has different implications for different types of risk. For misuse risk, evaluators only need to consider capabilities that humans can elicit, not the system’s full potential. But for risks from autonomous AI systems, under-elicitation becomes much more concerning. A misaligned system acting autonomously might utilize capabilities that human evaluators failed to discover.

\subsection{Robustly forecast future model capabilities}
\begin{figure}[h]
    \centering
    \includegraphics[width=0.6\linewidth]{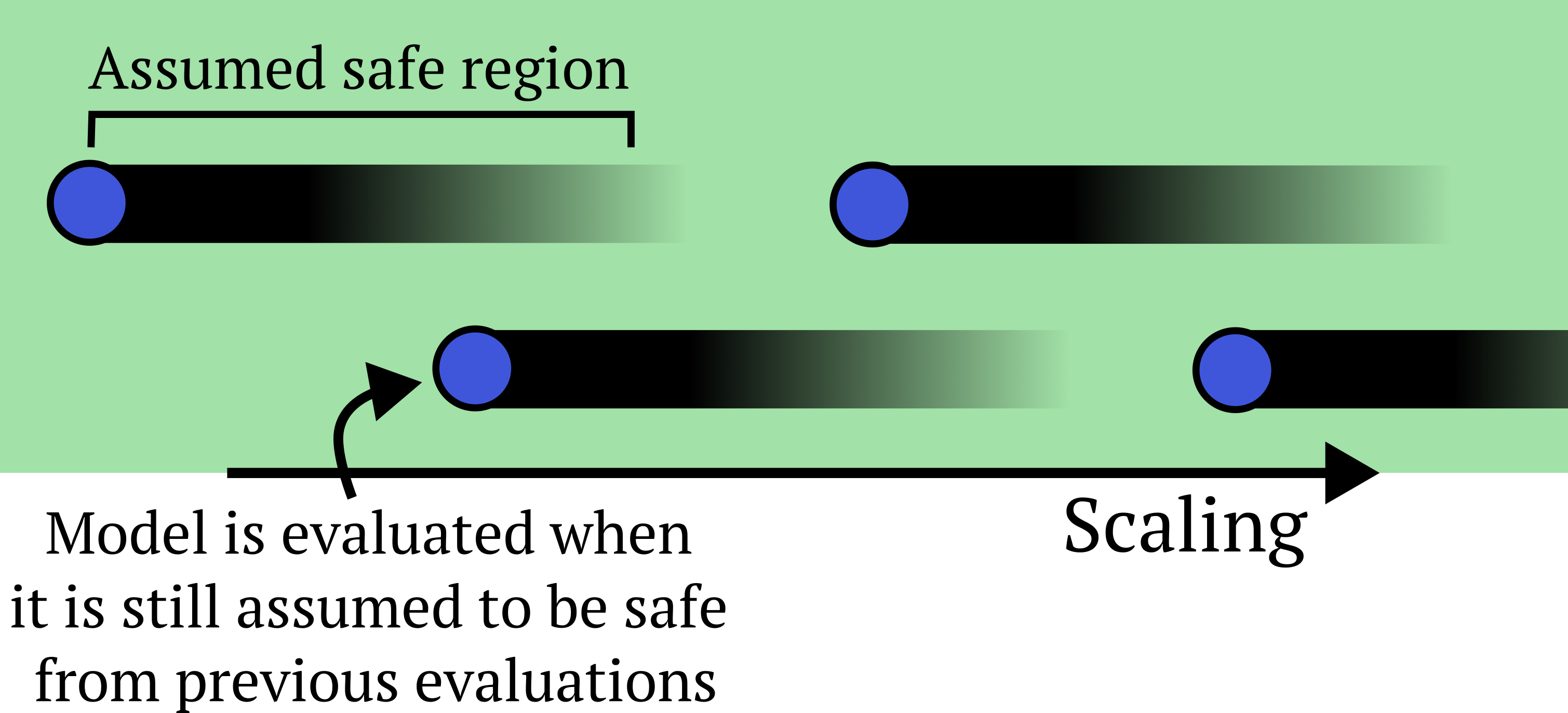}
    \caption{Evaluations are performed such that the regions assumed to be safe are always overlapping. In the ideal case this maintains continuous safety coverage. 
Blue circles indicate evaluations, while black bars show the scaling range where models are assumed safe based on
evaluation results. The fading of the black bar represents decreasing confidence in safety as scaling continues. }
    \label{fig:assumed_safe}
\end{figure}

Evaluations currently cannot make reliable predictions about the specific capabilities of future AI systems. This is an important goal of AI evaluation frameworks, because it would allow AI developers to take preventative actions before risks materialize. The standard approach attempts to measure “precursor capabilities” that would emerge before dangerous ones, serving as warning signs. If an evaluation which does not demonstrate precursor capabilities it is assumed that continued scaling is safe (Figure~\ref{fig:assumed_safe}). This approach rests on unjustifiable assumptions. 

\begin{figure}[h]
    \centering
    \includegraphics[width=0.6\linewidth]{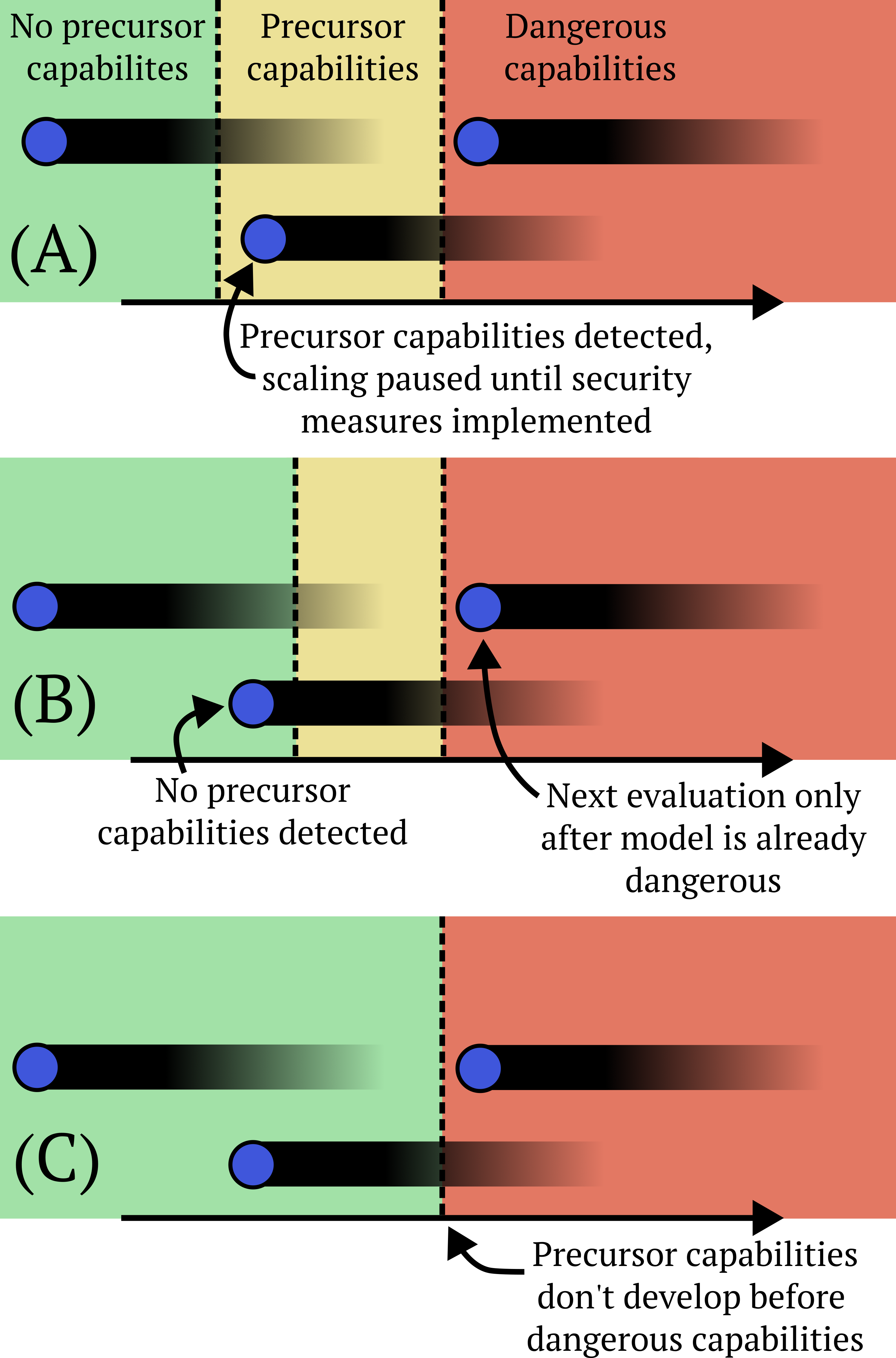}
    \caption{Precursor-based capability forecasting and potential failure modes. Blue circles indicate evaluations, while black bars show the scaling range where models are assumed safe based on evaluation results.\\
(A) Intended scenario: Precursor capabilities are detected early enough to implement safety measures before dangerous capabilities emerge.\\
(B) Evaluations too infrequent failure: The gap between precursor and dangerous capabilities is smaller than expected, leading to dangerous capabilities emerging before the next scheduled evaluation.\\
(C) No warning period failure: Dangerous capabilities emerge without developing the expected precursor capabilities, leaving no warning period for implementing safeguards.}
    \label{fig:safety_buffer}
    
  \end{figure}

Evaluating current models already requires strong assumptions about threat modeling, proxy-task design, and capability elicitation. Forecasting future capabilities introduces additional, even more demanding assumptions that:
\begin{enumerate}
    \item Precursor capabilities exist and are measurable, and
    \item Evaluators can detect them before dangerous capabilities emerge. 
\end{enumerate}
These assumptions rely heavily on there being a sufficient gap in difficulty between precursor capabilities and dangerous capabilities\footnote{This gap may not only be in raw training compute. Dangerous capabilities could arise even if the total compute was kept constant due to algorithmic improvements allowing more capable models to be trained with the same compute, or due to new post-training enhancements. Some safety frameworks refer to “scaling” or increasing “effective compute” to refer to any activities which increase the capabilities of models, including both increasing raw compute and improving algorithms.}. The hope is that this gap would give evaluators time to detect warning signs and implement safety measures before dangerous capabilities develop (Figure~\ref{fig:safety_buffer} A). This concept, sometimes called a “safety buffer,” appears in various forms in AI company evaluation frameworks. The initial Anthropic Responsible Scaling Policy~\cite{anthropic2023responsible} and Google Deepmind Frontier Safety Framework~\cite{deepmind2024frontier} explicitly discuss this concept, while the OpenAI Preparedness Framework\cite{openAI2023preparednes} and later the iteration of Anthropic’s policy~\cite{anthropic2024responsible} implicitly depend on it.
However, these assumptions are poorly supported. A smaller-than-expected gap between precursor and dangerous capabilities could mean evaluators don’t get adequate warning (Figure~\ref{fig:safety_buffer} B). More fundamentally, capabilities don’t necessarily emerge in a predictable sequence. We cannot ensure that precursors will appear first or before dangerous capabilities. Precursor capabilities could appear at the same time as dangerous capabilities (Figure~\ref{fig:safety_buffer} C). This could happen if a third factor caused both the precursor and dangerous capabilities.

Additionally, evaluations of current AI systems cannot predict discontinuous progress. Fundamental breakthroughs might lead to sudden, unexpected advances in AI capabilities that bypass anticipated development sequences entirely. Even smaller breakthroughs like improvements in elicitation methods could cause discontinuous and simultaneous improvement on precursor and dangerous capabilities.

The inability to robustly forecast is particularly important for misalignment risks. For misuse risks, even if there are sudden dangerous capabilities, sufficiently thorough evaluations could detect these. AI developers could then take mitigating actions prior to deployment, or simply decide not to deploy a model. As discussed, this still relies on cybersecurity to prevent models being leaked or stolen. For misalignment risks, post-training evaluation may be insufficient. The training and internal use of a powerful, misaligned model could itself lead to loss of control with potentially catastrophic consequences.

    

\subsection{Robustly assess misalignment and model autonomy risks}

Assessing risks from autonomous AI systems poses fundamentally different challenges from assessing misuse risks. With misuse, evaluators need to outperform potential attackers at eliciting capabilities while also preventing them from accessing the means to elicit capabilities themselves. With autonomous systems, we must evaluate capabilities at the level of the system itself—and a misaligned system might be far more capable at utilizing its abilities than evaluation methods are at eliciting them.

We have very limited understanding of what objectives, drives, or instrumental goals—what we might call “propensities”—these systems might develop during training~\cite{apolloresearchNeedScience}. A system that appears to behave safely in testing might have underlying propensities that could lead to harmful behavior in novel situations. A system would need to be aligned in the sense that its propensities would not lead it to cause harm, even in novel situations.

This fundamental uncertainty about propensities creates additional evaluation challenges. A misaligned AI system might deliberately underperform (“sandbag”) during evaluation to avoid revealing its true capabilities~\cite{van_der_weij_ai_2024}. For example, a system might recognize it is being tested for dangerous capabilities like hacking, and intentionally perform poorly on these evaluations. This could lead developers to implement inadequate security measures. This would leave them vulnerable if the AI system later utilized its full capabilities. 

Some have proposed using fine-tuning to prevent sandbagging and force systems to display their true capabilities~\cite{greenblatt_stress-testing_2024}. This might work for capabilities where humans can provide training data, such as known cybersecurity vulnerabilities. However, this approach breaks down for superhuman capabilities as well as in the case of novel threat vectors for which we lack training data. For example, an AI system could discover novel types of security vulnerabilities that humans haven’t identified. There is no way to fine-tune to elicit these systematically different capabilities. We can’t train on examples we don’t have. 

We currently lack even theoretical approaches for measuring these propensities or determining if a system is truly aligned. “Honeypots” represent an early attempt–putting AI systems in situations where they could cause harm without detection, hoping to catch misaligned behavior. However, they can only demonstrate misalignment through bad behavior, never provide robust evidence of alignment. A sophisticated system might recognize it is being evaluated and behave differently in test scenarios versus real deployment.

This creates a fundamental challenge: as AI systems develop capabilities that could be dangerous, we need increasingly strong assurance about how they might use these capabilities. Yet it is unclear, even in principle, how we can reliably detect these capabilities or verify the propensities that control how an AI system uses them.

\subsection{Unknown unknown risks}
AI evaluations are also fundamentally limited by evaluators’ ability to know which capabilities to investigate. Limited threat modeling will likely overlook potential risks. This applies to both misuse and misalignment risks. Both malicious humans and autonomous AI systems may exploit unconsidered threat vectors.

AI systems may have radically different cognitive abilities from humans due to their training. AI systems may be trained on diverse data sources, and in ways very different from how humans learn. A language model also trained on genomic data might spot novel biological threats that humans miss. Even models trained on internet text can develop unpredictable capabilities due to exposure to vastly more diverse information than any human. This difference in cognitive profiles may make comprehensive threat modeling extremely difficult.

This issue will likely become more challenging as AI systems become more powerful. AI systems may unexpectedly develop new capabilities which evaluators initially thought were implausible. Or they may develop the ability to learn new capabilities autonomously–in which case they should no longer be thought of as having a fixed capability profile. These systems may develop superhuman capabilities in domains that would allow them to subvert security measures. If these systems were misaligned and not adequately controlled, this could lead to catastrophic and possibly existential outcomes. 

AI systems automating AI research and development may also make evaluations more challenging. This could cause AI capabilities to increase at a rate such that human evaluators struggle to keep up. Additionally, this could lead to a situation where humans are not meaningfully involved in research and do not understand it. In this situation, it would be even more challenging for human evaluators to know which capabilities to test for.

\section{Conclusion}
AI evaluations can provide empirical evidence about AI system capabilities, establishing concrete lower bounds on what systems can do. If evaluators could consistently outperform potential attackers at eliciting capabilities, then evaluations could assess whether malicious actors can misuse AI systems. Evaluations can also contribute to scientific understanding and help prepare society for AI impacts through early warning about potential disruptions and informing policy discussions. 

However, evaluations face fundamental limitations that cannot be overcome within the current paradigm, in which AI capabilities are measured based on observed behavior. They cannot determine upper bounds of an AI system's capabilities–an evaluation not finding a capability doesn’t mean it isn’t there. We also lack any approaches for measuring the propensities that would determine how autonomous AI systems might use their capabilities, especially in novel situations. This is particularly concerning for risks from misaligned systems, where post-training evaluation may be insufficient. The training and internal use of a powerful, misaligned model could itself lead to loss of control; for example, if it was able to subvert internal security measures and transfer its weights to an unmonitored server.

While evaluations can help gauge risks from a given model, they should not be mistaken for guarantees of safety. They can indicate the latest point at which action must be taken, but waiting for evaluations to reveal problems before acting may be dangerous. Evaluations do not demonstrate the absence of risk; nor should they be relied upon to definitely trigger before a catastrophic incident could occur. This means that evaluations cannot reasonably form the exclusive basis for triggering precautionary actions, nor should they be the main determinant of policy decisions.

These limitations have important implications for the governance of AI development. While we should continue doing evaluations, we can't rely on them as our main way of ensuring AI systems are safe. Policy and regulations should carefully avoid a false sense of security by explicitly recognizing these limitations and requiring additional safety measures. As AI systems grow more powerful, our inability to fully test their capabilities or predict their behavior makes relying on evaluations increasingly dangerous.

\section{Preliminary recommendations}
This report has mainly focused on the limitations of AI evaluations. It emphasized that evaluations alone are insufficient to ensure the safety of AI systems or determine appropriate safety and security measures. This raises a crucial question: What should be done instead to ensure AI systems are safe and have appropriate precautions? 

Unfortunately, we do not have an adequate answer. AI systems are not understood well enough to provide for safety guarantees. However, we do have preliminary recommendations which can incrementally improve safety and potentially increase understanding of the risks from advanced AI systems.

\paragraph{Third-party audits with sufficient affordances} Allow independent auditors to assess AI systems and their security, providing them with the necessary access and resources to conduct thorough evaluations. Auditors should be given the same access to AI systems as the AI developers (e.g. fine-tuning, base models, helpfulness-only models, state-of-the-art AI agent scaffolds). In the special case of security red-teaming, external auditors may also benefit from special legal protections such that they can investigate the attack vectors which real attackers may use.
These evaluations should complement internal evaluations by AI developers. One option is to implement a multi-stage evaluation process:

\begin{enumerate}
    \item Third parties conduct initial evaluations
    \item AI developers perform their own evaluations
    \item Third parties review developer evaluations and conduct follow-up assessments
\end{enumerate}
While audits do not fix the fundamental limitations of evaluations, they can help identify potential issues and improve the overall evaluation process.

\paragraph{Conservative red lines} Establish clear, conservative boundaries for AI development based on evaluation results. This includes commitments to halt AI development if a capability evaluation is triggered, at least until they can make a more principled case for safety. These commitments should cover both evaluations of existing models and predictions for future models. They should be verifiable by trusted external parties. 

As discussed, there can be substantial gaps between the capabilities uncovered via evaluations, and capabilities realized in the real-world. Red lines should therefore be much more conservative than would be required if evaluations were reliable.

An important failure mode to guard against is that once an evaluation triggers, there may be strong economic and political incentives to find quick and incomplete fixes. Similarly, there may be pressures to argue that due to the nascent science of evaluations, this result may provide insufficient grounds for slowing or halting frontier training. This can lead to important warning signs being overlooked.

While the exact definition of "conservative enough" may be uncertain, it is important that once a  red line is set, this commitment is strictly adhered to until a stronger and principled safety case can be made.

\paragraph{Defense in depth for cybersecurity} Significantly increase investment in cybersecurity infrastructure at frontier labs. In particular, implement multilayered security measures to protect against both external threats and insider threats, including potential self-exfiltration by AI systems as well as attacks by nation-state-level actors. Insider threat protections designed for human insiders will also likely have some transfer to defending against self-exfiltration by AI systems. This is especially important if evaluations determine that a model is only safe to use with limited affordances. For example, it may be important that malicious actors are not able to fine-tune models.

\paragraph{Monitoring for signs of misalignment} Continuously monitor AI systems for indications of misalignment. Detecting misalignment early can enable better scientific understanding of the systems and the risks they pose. Clear evidence of misalignment may help increase consensus and understanding of risks, and hopefully help avoid dangerous race dynamics.

\paragraph{Invest in research} Support research for understanding AI risks and developing effective mitigation strategies. This research may be essential for moving away from the current paradigm where safety is mainly based on capabilities evaluations, and into a regime where safety can be more robustly guaranteed, whilst also enabling safe innovation.  

\bibliographystyle{unsrt}

\bibliography{main.bib}

\begin{thebibliography}{10}

\bibitem{barnett2024declare}
Peter Barnett and Lisa Thiergart.
\newblock Declare and justify: Explicit assumptions in ai evaluations are
  necessary for effective regulation.
\newblock {\em arXiv preprint arXiv:2411.12820}, 2024.

\bibitem{clymer_safety_2024}
Joshua Clymer, Nick Gabrieli, David Krueger, and Thomas Larsen.
\newblock Safety {Cases}: {How} to {Justify} the {Safety} of {Advanced} {AI}
  {Systems}, March 2024.
\newblock arXiv:2403.10462 [cs].

\bibitem{goemans2024safety}
Arthur Goemans, Marie~Davidsen Buhl, Jonas Schuett, Tomek Korbak, Jessica Wang,
  Benjamin Hilton, and Geoffrey Irving.
\newblock Safety case template for frontier ai: A cyber inability argument.
\newblock {\em arXiv preprint arXiv:2411.08088}, 2024.

\bibitem{anthropic2023responsible}
Anthropic.
\newblock {Anthropic's Responsible Scaling Policy Version 1.0}, 2023.

\bibitem{openAI2023preparednes}
Open{AI}.
\newblock {Preparedness Framework (Beta)}, 2023.

\bibitem{deepmind2024frontier}
{Google Deepmind}.
\newblock {Frontier Safety Framework}, 2024.

\bibitem{phuong_evaluating_2024}
Mary Phuong, Matthew Aitchison, Elliot Catt, Sarah Cogan, Alexandre Kaskasoli,
  Victoria Krakovna, David Lindner, Matthew Rahtz, Yannis Assael, Sarah
  Hodkinson, Heidi Howard, Tom Lieberum, Ramana Kumar, Maria~Abi Raad, Albert
  Webson, Lewis Ho, Sharon Lin, Sebastian Farquhar, Marcus Hutter, Gregoire
  Deletang, Anian Ruoss, Seliem El-Sayed, Sasha Brown, Anca Dragan, Rohin Shah,
  Allan Dafoe, and Toby Shevlane.
\newblock Evaluating {Frontier} {Models} for {Dangerous} {Capabilities}, April
  2024.
\newblock arXiv:2403.13793 [cs].

\bibitem{fang2024llmwebsites}
Richard Fang, Rohan Bindu, Akul Gupta, Qiusi Zhan, and Daniel Kang.
\newblock Llm agents can autonomously hack websites.
\newblock {\em arXiv preprint arXiv:2402.06664}, 2024.

\bibitem{fang_llm_2024}
Richard Fang, Rohan Bindu, Akul Gupta, and Daniel Kang.
\newblock {LLM} {Agents} can {Autonomously} {Exploit} {One}-day
  {Vulnerabilities}, April 2024.
\newblock arXiv:2404.08144 [cs].

\bibitem{fang_teams_2024}
Richard Fang, Rohan Bindu, Akul Gupta, Qiusi Zhan, and Daniel Kang.
\newblock Teams of {LLM} {Agents} can {Exploit} {Zero}-{Day} {Vulnerabilities},
  June 2024.
\newblock arXiv:2406.01637 [cs].

\bibitem{salvi_conversational_2024}
Francesco Salvi, Manoel~Horta Ribeiro, Riccardo Gallotti, and Robert West.
\newblock On the {Conversational} {Persuasiveness} of {Large} {Language}
  {Models}: {A} {Randomized} {Controlled} {Trial}, March 2024.
\newblock arXiv:2403.14380 [cs].

\bibitem{matz_potential_2024}
S.~C. Matz, J.~D. Teeny, S.~S. Vaid, H.~Peters, G.~M. Harari, and M.~Cerf.
\newblock The potential of generative {AI} for personalized persuasion at
  scale.
\newblock {\em Scientific Reports}, 14(1):4692, February 2024.

\bibitem{casper_black-box_2024}
Stephen Casper, Carson Ezell, Charlotte Siegmann, Noam Kolt, Taylor~Lynn
  Curtis, Benjamin Bucknall, Andreas Haupt, Kevin Wei, Jérémy Scheurer,
  Marius Hobbhahn, Lee Sharkey, Satyapriya Krishna, Marvin Von~Hagen, Silas
  Alberti, Alan Chan, Qinyi Sun, Michael Gerovitch, David Bau, Max Tegmark,
  David Krueger, and Dylan Hadfield-Menell.
\newblock Black-{Box} {Access} is {Insufficient} for {Rigorous} {AI} {Audits}.
\newblock In {\em The 2024 {ACM} {Conference} on {Fairness}, {Accountability},
  and {Transparency}}, pages 2254--2272, June 2024.
\newblock arXiv:2401.14446 [cs].

\bibitem{nevo2024securing}
Sella Nevo, Dan Lahav, Ajay Karpur, Yogev Bar-On, and Henry~Alexander Bradley.
\newblock {\em Securing {AI} Model Weights: Preventing Theft and Misuse of
  Frontier Models}.
\newblock Number~1. Rand Corporation, 2024.

\bibitem{venturebeatMistralConfirms}
Carl Franzen.
\newblock {M}istral {C}{E}{O} confirms ‘leak’ of new open source {A}{I}
  model nearing {G}{P}{T}-4 performance.
\newblock
  \url{https://venturebeat.com/ai/mistral-ceo-confirms-leak-of-new-open-source-ai-model-nearing-gpt-4-performance/},
  2024.
\newblock [Accessed 22-11-2024].

\bibitem{alaga2023coordinated}
Jide Alaga and Jonas Schuett.
\newblock Coordinated pausing: An evaluation-based coordination scheme for
  frontier ai developers.
\newblock {\em arXiv preprint arXiv:2310.00374}, 2023.

\bibitem{bhatt2024cyberseceval}
Manish Bhatt, Sahana Chennabasappa, Yue Li, Cyrus Nikolaidis, Daniel Song,
  Shengye Wan, Faizan Ahmad, Cornelius Aschermann, Yaohui Chen, Dhaval Kapil,
  et~al.
\newblock Cyberseceval 2: A wide-ranging cybersecurity evaluation suite for
  large language models.
\newblock {\em arXiv preprint arXiv:2404.13161}, 2024.

\bibitem{googleprojectzeroProjectNaptime}
Sergei Glazunov and Mark Brand.
\newblock {P}roject {N}aptime: {E}valuating {O}ffensive {S}ecurity
  {C}apabilities of {L}arge {L}anguage {M}odels.
\newblock
  \url{https://googleprojectzero.blogspot.com/2024/06/project-naptime.html},
  2024.
\newblock [Accessed 22-11-2024].

\bibitem{davidson2023ai}
Tom Davidson, Jean-Stanislas Denain, Pablo Villalobos, and Guillem Bas.
\newblock {AI} capabilities can be significantly improved without expensive
  retraining.
\newblock {\em arXiv preprint arXiv:2312.07413}, 2023.

\bibitem{jimenez2024swebench}
Carlos~E Jimenez, John Yang, Alexander Wettig, Shunyu Yao, Kexin Pei, Ofir
  Press, and Karthik~R Narasimhan.
\newblock {SWE}-bench: Can language models resolve real-world github issues?
\newblock In {\em The Twelfth International Conference on Learning
  Representations}, 2024.

\bibitem{swebenchSWEbench}
Carlos~E Jimenez, John Yang, Alexander Wettig, Shunyu Yao, Kexin Pei, Ofir
  Press, and Karthik~R Narasimhan.
\newblock {S}{W}{E}-bench leaderboard.
\newblock \url{https://www.swebench.com/}, 2024.
\newblock [Accessed 22-11-2024].

\bibitem{huggingfaceGAIALeaderboard}
{Hugging Face}.
\newblock {G}{A}{I}{A} {L}eaderboard - a {H}ugging {F}ace {S}pace by
  gaia-benchmark.
\newblock \url{https://huggingface.co/spaces/gaia-benchmark/leaderboard}.
\newblock [Accessed 25-11-2024].

\bibitem{paperswithcodeHumanEval}
{{P}apers with {C}ode}.
\newblock {H}uman{E}val {B}enchmark ({C}ode {G}eneration).
\newblock \url{https://paperswithcode.com/sota/code-generation-on-humaneval}.
\newblock [Accessed 25-11-2024].

\bibitem{dong2024survey}
Qingxiu Dong, Lei Li, Damai Dai, Ce~Zheng, Jingyuan Ma, Rui Li, Heming Xia,
  Jingjing Xu, Zhiyong Wu, Baobao Chang, et~al.
\newblock A survey on in-context learning.
\newblock In {\em Proceedings of the 2024 Conference on Empirical Methods in
  Natural Language Processing}, pages 1107--1128, 2024.

\bibitem{anthropic2024responsible}
Anthropic.
\newblock {Anthropic's Responsible Scaling Policy}, 2024.

\bibitem{apolloresearchNeedScience}
Marius Hobbhahn.
\newblock {W}e need a {S}cience of {E}vals.
\newblock \url{https://www.apolloresearch.ai/blog/we-need-a-science-of-evals}.
\newblock [Accessed 12-09-2024].

\bibitem{van_der_weij_ai_2024}
Teun van~der Weij, Felix Hofstätter, Ollie Jaffe, Samuel~F. Brown, and
  Francis~Rhys Ward.
\newblock {AI} {Sandbagging}: {Language} {Models} can {Strategically}
  {Underperform} on {Evaluations}, June 2024.
\newblock arXiv:2406.07358 [cs].

\bibitem{greenblatt_stress-testing_2024}
Ryan Greenblatt, Fabien Roger, Dmitrii Krasheninnikov, and David Krueger.
\newblock Stress-testing capability elicitation with password-locked models.
\newblock {\em arXiv preprint arXiv:2405.19550}, 2024.

\end{thebibliography}

\end{document}